\documentclass[a4paper]{article}
\usepackage{INTERSPEECH2021}

\graphicspath{ {./images/} }
\usepackage{adjustbox}
\usepackage{graphicx}
\usepackage{multirow}
\usepackage{textcomp}

\usepackage[table,xcdraw]{xcolor}
\usepackage{url}
\usepackage[colorlinks,citecolor=blue,urlcolor=blue,pdfsubject={LaTeX}]{hyperref}
\PassOptionsToPackage{hyphens}{url}
\usepackage[hyphenbreaks]{breakurl}
\usepackage[keeplastbox]{flushend}

\def \sharath [#1]{\textcolor{blue}{SA: #1}}  
\def \pranav [#1]{\textcolor{red}{p: #1}}
\def \nagaraj [#1]{\textcolor{magenta}{NA: #1}}
\def \arun [#1]{\textcolor{green}{AB: #1}}

\title{Non-native English lexicon creation for bilingual speech synthesis}
\name{Arun Baby, Pranav Jawale, Saranya Vinnaitherthan, Sumukh Badam, \\Nagaraj Adiga, Sharath Adavanne}

\address{
  Zapr Media Labs (Red Brick Lane Marketing Solutions Pvt. Ltd.), India}
\email{arun.baby@zapr.in}

\begin{document}

\maketitle
\begin{abstract}

Bilingual English speakers speak English as one of their languages. Their English is of a non-native kind, and their conversations are of a code-mixed fashion. The intelligibility of a bilingual text-to-speech (TTS) system for such non-native English speakers depends on a lexicon that captures the phoneme sequence used by non-native speakers. However, due to the lack of non-native English lexicon,  existing bilingual TTS systems employ native English lexicons that are widely available, in addition to their native language lexicon. Due to the inconsistency between the non-native English pronunciation in the audio and native English lexicon in the text, the intelligibility of synthesized speech in such TTS systems is significantly reduced.

This paper is motivated by the knowledge that the native language of the speaker highly influences non-native English pronunciation. We propose a generic approach to obtain rules based on letter to phoneme alignment to map native English lexicon to their non-native version. The effectiveness of such mapping is studied by comparing bilingual (Indian English and Hindi) TTS systems trained with and without the proposed rules. The subjective evaluation shows that the bilingual TTS system trained with the proposed non-native English lexicon rules obtains a 6\% absolute improvement in preference.

\end{abstract}

\section{Introduction}
Developing a bilingual text-to-speech (TTS) system~\cite{chu2003microsoft} is necessary for countries like India where the majority of the population speak more than one language. Generally, this population speaks their native language as the first and English as their second language. The pronunciation of English words by a non-native speaker is strongly influenced by their native language and is most often different from the native English pronunciation~\cite{lee2018learning}. Indian languages, which have a high grapheme to phoneme correlation (phonemic language), derive pronunciation directly from the spellings of the word. On the contrary, English is an alphabetic and highly non-phonemic language. Hence native phonemic language speakers whose pronunciation is influenced by the spelling of the word often pronounce English words differently from native English speakers. This mispronunciation is further enhanced for native speakers from languages whose phonemes are different from the English language. These speakers generally replace the English phoneme with the closest phoneme in their native language. Given these challenges, building a TTS system for such non-native English bilingual speakers requires a lexicon that handles the influence of the first language on the native English lexicon. However, due to the lack of availability of such a non-native English lexicon~\cite{kumar2007building}, existing bilingual TTS systems employ widely available native English lexicon, in addition to their native language lexicon, which results in reduced intelligibility and heavily accented synthesized speech. 

Yarra~\emph{et al.}~\cite{yarra2019indic} proposed to collect Indian English lexicon from the mispronunciations of Indian speakers recording sentences from the TIMIT database. Multiple phonemes- and letter-specific context rules were manually identified through observing pronunciation variations between Indian and original TIMIT speakers. The proposed lexicon was evaluated on the speech recognition task and shown to improve the overall recognition. However, collecting such parallel data for a large vocabulary and manually finding the mispronunciation between the two is a tedious and expensive task. Anju~\emph{et al.}~\cite{Thomas2018} proposed a transliteration-based method for developing speech synthesizers for Bilingual Indian English. A mapping was obtained between native English CMU dictionary phonemes and the Indian common phone Label Set (CLS)~\cite{ramani2013common} shared across the Indian languages. The stress markers of the CMU phoneset were removed before this mapping. Further, the words not part of the CMU dictionary were first transliterated to a phonemic Indian language. The CLS phoneme sequence was obtained using the unified parser~\cite{nlp:tsd16conf} grapheme to phoneme (G2P) model. %[?]A version of this lexicon is employed as the baseline (Model 2 in Section~\ref{ssec:Model2}) in this paper. 
In~\cite{mahanta2016text}, the phoneme sequence of the CMU dictionary was manually corrected for the Assamese language accent to develop an Assamese English TTS system. However, the manual effort of editing the lexicon is an expensive and time-consuming operation. In~\cite{saikia2016generating}, a sequence labelling approach is employed to generate a pronunciation dictionary using Conditional Random Fields (CRFs). However, this method needs a substantial parallel corpus of phone sequence mapping between native and non-native lexicon to train the CRF network.

In this paper, we propose a generic framework to obtain the rules for mapping the phone sequence of a native English lexicon to a non-native one. Specifically, the framework aims to derive non-native English pronunciation for speakers from native languages that follow phonemic orthography. Although we study the framework on native Hindi language speaker,  the framework itself should be adaptable to speakers of other phonemic languages. As the first step, we identify a subset of highly frequent English words. For these words, a three-way alignment is obtained between a) the English letter sequence, b) the CMU phoneme sequence from native English CMU dictionary, and c) the CLS phoneme sequence obtained using unified parser~\cite{nlp:tsd16conf} from the (manually curated) transliterated version of the English word. Repeating patterns of the aligned triplets of letter-CMU-CLS phonemes that produce mispronunciations are manually identified. After that, rules are devised which, wherever relevant, will substitute an original CMU phoneme in a word pronunciation with a new phoneme that produces the correct non-native English pronunciation. These rules are applied on the entire native English lexicon to obtain the corresponding non-Native English lexicon. We show that the proposed framework provides better non-native English pronunciation than the existing frameworks through subjective listening tests.

\section{Proposed approach}
\label{sec:proposed}
The proposed approach is studied on a bilingual Indian English-Hindi dataset~\cite{babycbblr2016}. This dataset contains monolingual recordings in English and Hindi from the same Male speaker (More about the dataset in Section~\ref{ssec:data}). The English transcripts are in Roman script, while Hindi is in Devanagari script. The pronunciations for English words were obtained from g2p\_en\footnote{https://pypi.org/project/g2p-en/} G2P model which uses the 39 CMU phoneset. Similarly, the Unified Parser~\cite{nlp:tsd16conf} was employed to obtain the pronunciations for the Hindi words, that used the 59 CLS phonemes, amounting to 98 phonemes in total for the bilingual dataset. Further, motivated by the works of~\cite{Thomas2018}, we merged acoustically similar CMU and CLS phonemes~\cite{vijayarajsolomon2016exploiting}. This merging is intended to address non-native English speakers substituting similar phonemes from their native language in place for native English phonemes. Additionally, reduced phoneset results in increased training data per phoneme, and consequently better phoneme modelling. This reduced phoneset has 73 phonemes and is referred to \mbox{``EngHinCommon"} hereafter. While merging, care was taken not to merge phonemes whose realization is audibly distinct across Hindi and English languages. The merged CMU and CLS phonemes are listed in Table~\ref{img:EngHinCommon}. 

As the baseline for the proposed approach, we trained separate bilingual TTS systems with the above 98 CMU and CLS phoneset (Model 1 in Section~\ref{ssec:Exp}) and the reduced EngHinCommon phoneset (Model 2 in Section~\ref{ssec:Exp}). The results showed that the quality of synthesis of Hindi words remained consistent between the two models. However, for the English words, contrary to the motivation of employing the reduced phoneset, the non-native English pronunciation improved only for a few English words compared to Model 1. More details of these experiments are discussed in Section~\ref{sec:results}. As discussed earlier, a non-native English speaker (whose native language follows phonemic orthography) takes a cue from the spelling of the word for pronunciation. For example, The CMU phone sequence for word /CITED/ is \mbox{/S AY T \textbf{AH} D/}, whereas a phonemic language speaker, takes a cue from the spelling and pronounces it as \mbox{/S AY T \textbf{EH} D/}. However, similar to \mbox{EngHinCommon} phoneset creation, a universal mapping of all /AH/ to /EH/ will create more problems than it solves. We need a nuanced approach that goes beyond one-to-one phoneme mapping. Motivated by this, in the following section, we explain the proposed approach to map phonemes based on additional information such as English letter identity, position within a syllable, and letter context.

\subsection{Selection of words}
The bilingual dataset has a vocabulary of about 7 k words, of these, around 6.5 k words are English words that are not proper nouns of Indian origin. A subset of 2 k words from 6.5 k words of the dataset is chosen based on their being present in the top 10 k most frequent words in an independently obtained Indian newspaper text archive. It is crucial to select a large number of words to arrive at accurate and exhaustive rules.

This paper does not study proper nouns of Indian origin because the native English G2P models fail on these words, making it impossible to recover from these errors even after using a rule-based phoneme mapping. The best way to derive pronunciation for such Indian origin proper nouns is to transliterate them into a phonemic language script and obtain the phoneme sequence using a G2P model like Unified Parser~\cite{nlp:tsd16conf}.

%\pranav[change figure 1 fron new screenshot]{DONE!}

\begin{table}[t]
    % \centering
    \centerline{\includegraphics[width=\columnwidth,height=11cm,keepaspectratio]{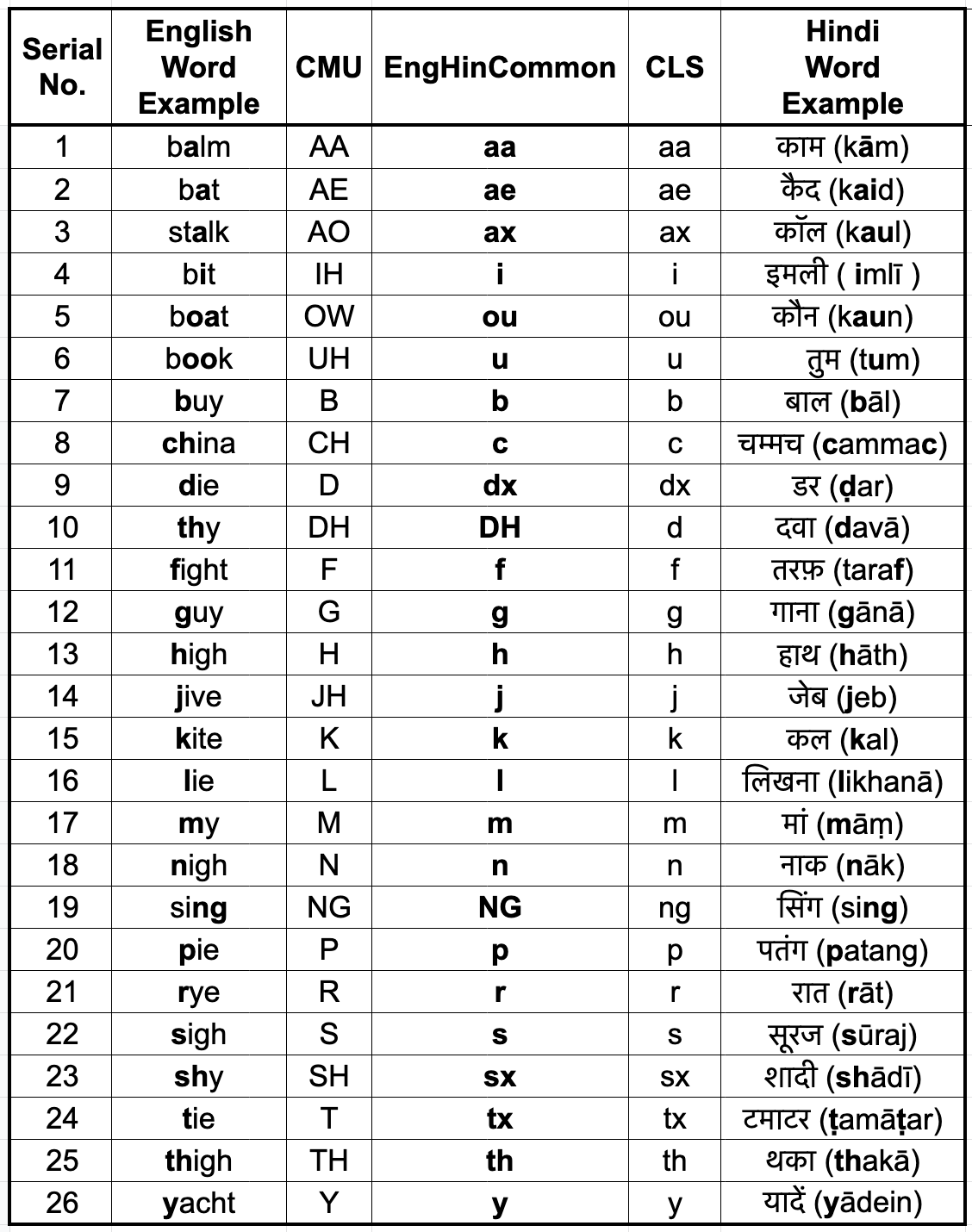}}
    \caption{Subset of EngHinCommon phonemes that are obtained by merging CMU and CLS phoneset. Rest of the CMU and CLS phonemes are used as unique phonemes of \mbox{EngHinCommon.}}
    \label{img:EngHinCommon}
\end{table}

% \begin{figure}[t]
%     % \centering
%     \centerline{\includegraphics[width=\columnwidth,height=11cm,keepaspectratio]{images/EngHinCommon_Subset.png}}
%     \caption{Subset of EngHinCommon phonemes that are obtained by merging CMU and CLS phoneset. Rest of the CMU and CLS phonemes are used as unique phonemes of \mbox{EngHinCommon.}}
%     \label{img:EngHinCommon}
% \end{figure}

\subsection{Process of deriving rules}
\label{ssec:processofrules}
\begin{figure}[t]
    \centering
    \centerline{\includegraphics[width=\columnwidth, keepaspectratio]{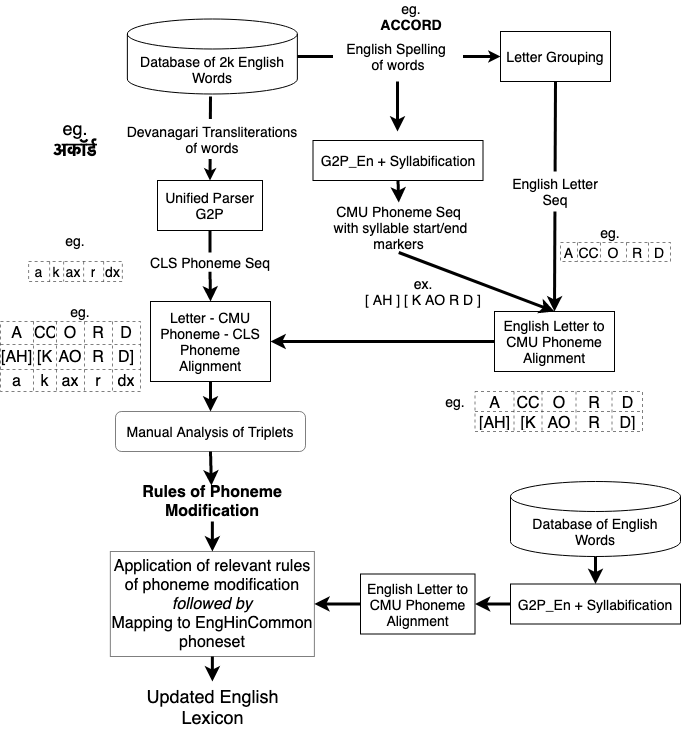}}
    \caption{Block diagram showing the process of deriving rules and the creation of non-native English lexicon.}
    \label{img:rules_derivation}
\end{figure}

The process of deriving phoneme mapping rules from the above subset of 2 k English words is described in Figure~\ref{img:rules_derivation}. To obtain the CLS phoneme sequence for these English words as pronounced by native Hindi speakers we used an online English-Hindi Dictionary\footnote{https://dict.hinkhoj.com/english-to-hindi/}, which has Devanagari transliteration for several of these words. Four native Hindi speakers conducted a manual verification to match the transliteration with the actual pronunciation of native Hindi speakers. To avoid any bias toward the speaker in the training dataset, the verification of transliteration was done purely based on text. Finally, the CLS phoneme sequences were obtained from these transliterations using the Unified Parser~\cite{nlp:tsd16conf}. We employ the transliterations only for the creation of rules. Once the rules are created, they can be directly applied to any native English lexicon to obtain their non-native English versions without the requirement of any transliteration.

Next, we used an alignment algorithm (explained in section~\ref{ssec:alignment}) to align the English letters in the words, the CMU phonemes for the word, and the target CLS phonemes. While in a large number of cases, source CMU phonemes are exclusively aligned with a single CLS phoneme. In some cases, the mapping was not exclusive, i.e., the same CMU phoneme was mapped to different CLS phonemes which are acoustically distant (e.g. /AH/ $\rightarrow$ /a/ in some words and /AH/ $\rightarrow$ /o/ in other words). This ambiguity is the primary reason behind the inconsistency between native and non-native English lexicon. In such cases, we analyzed them further to devise rules for correcting the lexicons. These rules are based on knowledge of the source English letter, its position within a syllable, source CMU phone, and letter sequence context. We aim to propose rules that will reduce this phoneme mapping ambiguity and come up with conditions under which a particular target phoneme should be used (explained in the following sections). Finally, after correcting the lexicon using the proposed rules, all the ambiguous conditions discussed above are handled. After that, the remaining phonemes in the lexicon that are not yet modified with the proposed rules, are directly mapped to EngHinCommon phoneset as shown in Table~\ref{img:EngHinCommon}.

\subsection{Alignment Algorithm}
\label{ssec:alignment}
One of the key steps of the proposed approach is to align the phoneme sequences with the letters. If the number of letters in English words were precisely the same as the number of phonemes, their alignment would be straightforward. However, this is not always the case. Hence we made use of \textbf{a.} Letter Grouping Heuristics and \textbf{b}. Heuristics for zero distance letter-phoneme pairs. 

% \subsubsubsection{a. Letter Grouping}:
a. Letter Grouping:
The aim of this grouping is to obtain units of letter sequences that usually correspond to a single phoneme. The following consecutive letters are always considered as a single unit - \textit{ph}, \textit{ch}, \textit{ng}, \textit{sh}, \textit{th}, \textit{er}, \textit{ow}. Further, any duplicated consecutive non-vowel letters (Regex:  [$\wedge${\textit{aeiou}}]\{2\}), and sequence of a single vowel letter followed by a vowel letter or letter \textit{y} (Regex:  [{\textit{aeiou}}][{\textit{aeiouy}}]), are also considered as single unit.

% \subsubsubsection{b. Heuristics for zero distance letter-phoneme pairs}:
b. Heuristics for zero distance letter-phoneme pairs: Certain letter-phoneme pairs are considered as equivalent (eg. letter \textit{p} $\leftrightarrow$ CMU phoneme P,   letter unit \textit{ph} $\leftrightarrow$ CMU phoneme F, letter \textit{c} $\leftrightarrow$ CMU phoneme K, and so on). This equivalence information helps the dynamic programming-based alignment algorithm to output more accurate alignments with better causal correspondence between letters and phonemes. 

\begin{figure}[t]
    \centering
    \centerline{\includegraphics[width=\columnwidth, height=4.5cm, keepaspectratio]{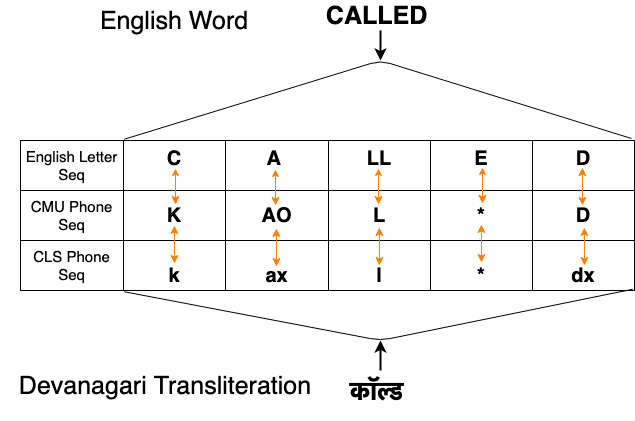}}
    \caption{The example illustrates the three-way alignment between English letters, CMU phone sequence, and CLS phone sequence obtained from the transliterated English word in Devanagari script}
    \label{img:3way}
\end{figure}

Similar equivalence relation between certain CMU and CLS phoneme pairs is used to finally get a three-way alignment between English letter sequence $\leftrightarrow$ CMU phoneme sequence $\leftrightarrow$ CLS Phoneme Sequence. A sample illustration of our alignment algorithm is shown in Figure~\ref{img:3way} for the word /CALLED/. Each column of the alignment in Figure~\ref{img:3way} is hereafter referred to as a triplet.

\begin{figure}[!b]
    \centering
    \centerline{\includegraphics[width=\columnwidth, height=5cm]{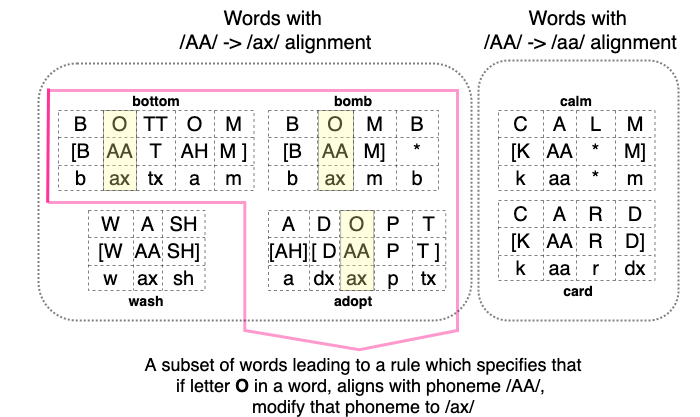}}
    \caption{Simplified illustration of triplet analysis used to derive the rule which states the condition under which we can modify phoneme /AA/ to phoneme /ax/}
    \label{img:/AA/ -> /ax/ rule derivation}
\end{figure}

\begin{table*}[ht]
    \centering
    \centerline{\includegraphics[width=18cm,keepaspectratio]{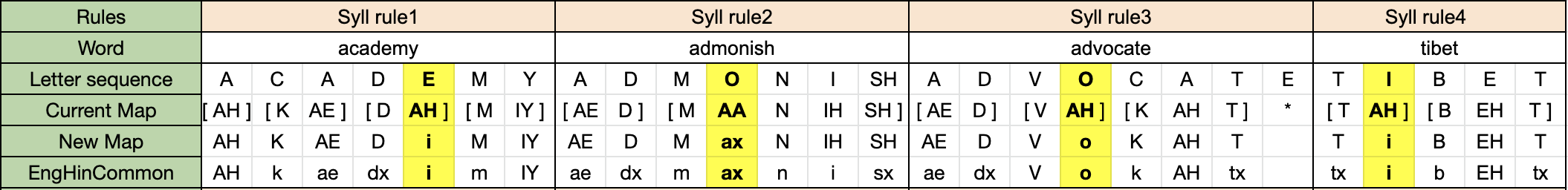}}
    \vspace{5pt}
    \centerline{\includegraphics[width=18cm,keepaspectratio]{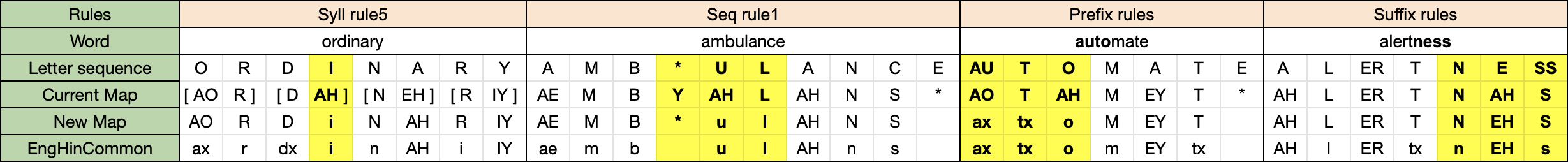}}
    \caption{Illustration of sample examples showing sub-set of rules and corrections at phone-level with triplet letter-CMU-CLS phoneme alignment. Modifications at the phoneme level are displayed in bold font.}
    \label{img:rules_egs}
\end{table*}

\subsection{Triplet Analysis}
\label{ssec:tripletAnalysis}
We analyse the triplets which contain non-exclusive CMU to CLS phone mapping to identify phoneme mapping patterns. For example, Figure~\ref{img:/AA/ -> /ax/ rule derivation} illustrates patterns where CMU phoneme /AA/ was aligned either with CLS phoneme /ax/ or with CLS phoneme /aa/ based on letter context present in triplets. We check if this knowledge about letters aligned with the CMU phoneme /AA/ can resolve this ambiguity. After going through a significant number of words resulting in such triplets, we develop a rule that applies to a large subset of words.

In some cases, to develop a specific rule, we have to refer to the position of the CMU phoneme within the syllable. The syllable boundaries are obtained using tsylb2 tool\footnote{https://www.nist.gov/itl/iad/mig/tools}. Syllable start and end boundaries within a CMU phone sequence are denoted using square brackets in Figure~\ref{img:rules_derivation} and Figure~\ref{img:/AA/ -> /ax/ rule derivation}.  Apart from syllable level corrections, in some cases, the rule is dependent upon the letter context (letters to the left and right) and whether that letter sequence occurs within the word or at suffix/prefix locations. From the triplet analysis, it was observed that for most high-frequency suffixes and prefixes (e.g. \textit{auto-}, \mbox{\textit{-ted}}, \textit{-less}, \textit{-ness}, \textit{-ment}), the Indian English speakers' pronunciation followed a consistent pattern. Therefore, additional rules for these suffixes and prefixes (40 rules) were identified from this analysis. A subset of these rules are depicted in Table~\ref{img:rules_egs} and Table~\ref{tab:rules}. 

In general, we have derived three kinds of rules -- \textbf{1.} Syllable level rules, \textbf{2.} Letter sequence rules, and \textbf{3.} Prefix/Suffix-based rules. Finally, using these rules the native English lexicon is mapped to non-native English lexicon for the Indian English - Hindi bilingual speaker as shown in Figure~\ref{img:rules_derivation}.  It is observed from these rules that most of the differences between native and Indian English pronunciation are among  vowels, this finding is consistent with prior work~\cite{vijayarajsolomon2016exploiting,maxwell2009acoustic}.

\begin{table}[t]
  \caption{Summary of proposed rules to map native English lexicon to corresponding non-native version. }
  \label{tab:rules}
  \centering
  \resizebox{0.49\textwidth}{!}{
    \begin{tabular}{c|c|c|c|c|c}
    %\hline
    % \multicolumn{5}{|c|}{\textbf{Rules}} & \textbf{CMUdict} \\ \hline
    \multicolumn{1}{c|}{\textbf{}} & \multicolumn{1}{c|}{\textbf{\begin{tabular}[c]{@{}c@{}}Source \\ English \\ Letter\end{tabular}}} & \multicolumn{1}{c|}{\textbf{\begin{tabular}[c]{@{}c@{}}Source \\ Phoneme \\ Sequence(CMU)\end{tabular}}} & \multicolumn{1}{c|}{\textbf{\begin{tabular}[c]{@{}c@{}}Target \\ Phoneme \\ Sequence(CLS)\end{tabular}}} & \textbf{\begin{tabular}[c]{@{}c@{}}Position \\ in \\ syllable\end{tabular}} & \textbf{\begin{tabular}[c]{@{}c@{}}Corrected\\ words in\\ CMUdict (\%)\end{tabular}} \\ \hline
    \textbf{Syll rule1} & E & AH & i & end & 1.9 \\ %\cline{1-1}
    \textbf{Syll rule2} & O & AA & ax & anywhere & 6.4 \\ %\cline{1-1}
    \textbf{Syll rule3} & O & AH & o & end & 2.7 \\ %\cline{1-1}
    \textbf{Syll rule4} & I & AH & i & end & 2.5 \\ %\cline{1-1}
    \textbf{Syll rule5} & A & EH & AH & end & 0.7 \\ \hline
    
    \textbf{Seq rule1} & (* U L) & (Y AH L) & (* u l) & anywhere & 0.2 \\ \hline
    \textbf{Prefix rules} & sub word & old sequence & new sequence & NA & 1 \\ %\cline{1-1}
    \textbf{Suffix rules} & sub word & old sequence & new sequence & NA & 19.8 \\ %\hline
    \end{tabular}
}
\end{table}

\section{Experiments}
\label{ssec:Exp}

\subsection{Dataset}
\label{ssec:data}
We selected a native Hindi Bilingual Male speaker from \mbox{IndicTTS}~\cite{babycbblr2016} database for our experimentation. This dataset consists of two parts, 9 hours each of monolingual Hindi and monolingual English dataset, amounting to ~18 hours of studio-quality recording by a professional voice-over artist. The original recordings are at a 48 kHz sampling rate. However, all the studies in this paper are performed at 16 kHz. The genre of the spoken text is fiction and children's stories. The transcripts for English is in Roman, and Hindi is in Devanagri script.

\subsection{TTS modeling}
A neural network-based TTS system is used for all the experiments. A neural TTS system is generally comprised of a front end and a vocoder. As the fronted, we use the Tacotron2 (v3) recipe of ESPnet~\cite{hayashi2019espnettts}, which is an auto-regressive-based sequence-to-sequence model with a location-sensitive and guided attention mechanism~\cite{shen2018natural}. The front end is trained with the phone sequence of the input text and the 80-band Mel spectrogram feature of the corresponding audio, computed with a 1024 point discrete Fourier transform and 256 sample hop-length. The front end is trained for 250 epochs with a batch size of 56 on 4 GPUs. The Mel spectrogram output of the front end is mapped to waveform using the parallel wavegan (PWG) vocoder~\cite{yamamoto2020parallel}. The PWG vocoder is a non-autoregressive variant of the WaveNet~\cite{oord2016wavenet} vocoder that has a significantly faster inference time. We use the publicly available implementation of PWG, whose code is accessible here\footnote{https://github.com/kan-bayashi/ParallelWaveGAN}.

\subsection{Systems}
We train three separate Tacotron2 models with different phonesets to evaluate the efficacy of the proposed method while keeping the vocoder fixed across all our experiments.

\subsubsection{Model 1} 
As the first baseline TTS model, we employ CMU phoneset (unstressed) for English data and CLS phoneset for Hindi. The unstressed version of CMU is used because Indian languages are syllable-timed and Indian speakers don't differentiate for different stress level~\cite{SIRSA2013393}. Here a total of 98 unique phonemes (39 CMU + 59 CLS) are used for training the Tacotron model and develop bilingual TTS. Refer to Section~\ref{sec:proposed} for more details.

\subsubsection{Model 2}
\label{ssec:Model2}
A second TTS model is trained to study the effect of using 73 phonemes EngHinCommon phoneset (Refer to Section~\ref{sec:proposed} for more details) on the pronunciation of English words. Apart from the number of phonemes, all other model parameters are identical to Model 1. 

\subsubsection{Model 3} 
Finally, we train a TTS model with the proposed phone mapping rules as described in Section~\ref{sec:proposed}. After that, we map the exclusively aligned CMU and CLS phones as discussed in Section~\ref{ssec:processofrules} to the EngHinCommon phoneset. The proposed rules correct around 40\% of the training data. Model 3 has an identical phoneset as Model 2.

% On top of this, a manual check and modification (around 14\% of the data had to be corrected manually) and modification of the whole 7 k training words are done beforehand to train this model. 

\subsection{Evaluations}
\label{ssec:Eval}
\begin{figure}[t]
    \centering
    \centerline{\includegraphics[width=\linewidth, height=4cm,keepaspectratio]{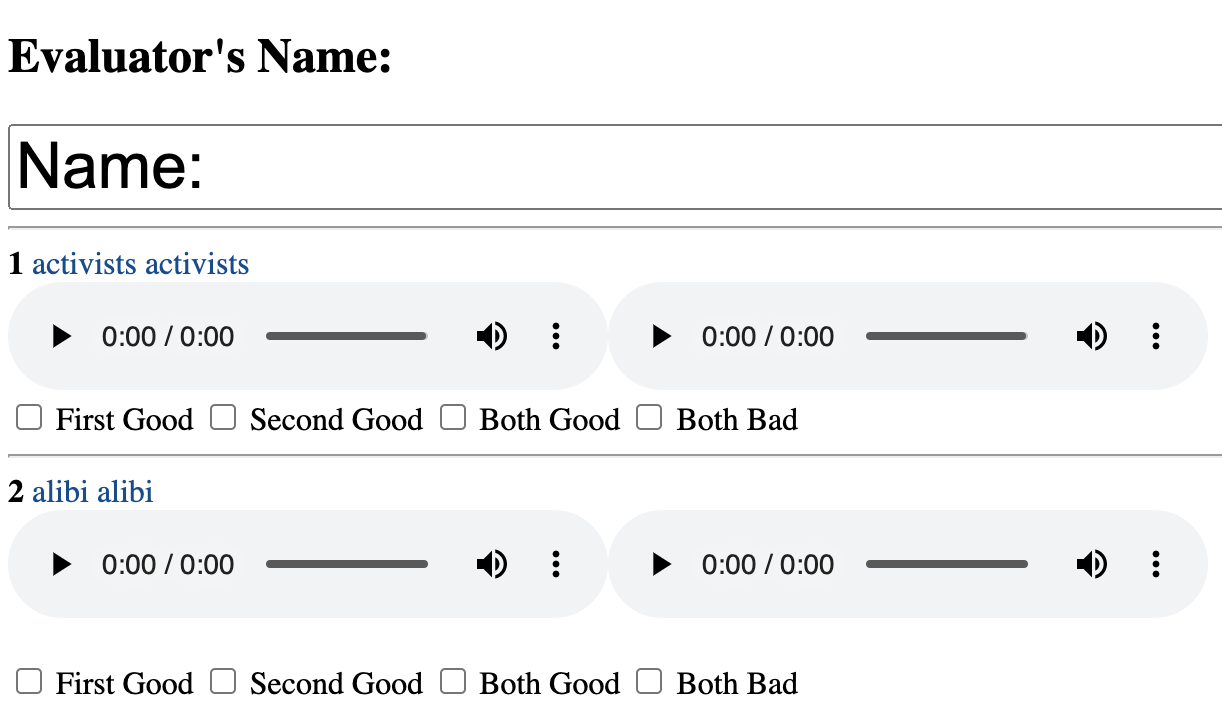}}
    \caption{Sample of subjective evaluation webpage employed to compare the pronunciation quality between two TTS models.}
    \label{img:eval_screen}
\end{figure}

Three different subjective evaluations are performed to assess the above TTS models. The evaluation is restricted to isolated words instead of a sentence to avoid any influence on the evaluator's rating by other words in the sentence. During each evaluation, a subset of English words are synthesized using two of the above models, and a preference test between the two is carried out as shown in Figure~\ref{img:eval_screen}. During the test, ten native Hindi speakers (also proficient in English) evaluate the synthesis of the models based on the closeness to the pronunciation of a bilingual Indian English-Hindi speaker. If the pronunciation of both the models were comparable, the listeners could choose the `Both Good' option, on the other hand, if both the pronunciations were bad, they could choose the `Both Bad' option. The synthesized recordings of the two models were shuffled and did not appear in the same order across the test. All evaluations are done with headphones to ensure clear perception, and the listeners were allowed to playback the audio any number of times. The synthesized recordings used for the subjective evaluations in our paper are available at \url{https://www.zapr.in/ssw2021/samples}.

\subsubsection{Evaluation 1}
As the first subjective evaluation, we compare Model 1 and Model 2 to study the effectiveness of the EngHinCommon phoneme mapping. The listeners rate a randomly chosen subset of 200 English words.

\subsubsection{Evaluation 2}
As the second subjective evaluation, Model 2 is compared against Model 3 to study the effectiveness of the proposed non-native English lexicon creation. For this evaluation, the listening test is carried out on three different subsets of 200 words: a) English words part of the training vocabulary, hereafter referred to as `Dict' words. b) English words not in the training vocabulary, but the rules have been applied (`Rules'). c) English words that were neither part of the training vocabulary nor modified by any rules (`OOR': Out-of-rules).

\subsubsection{Evaluation 3}
Finally, to complete the comparisons between the three models, Model 3 is compared with the baseline Model 1. Similar to Evaluation 2, three comparisons are made with Dict words, Rules words, and OOR words, respectively.

\label{sec:results}
\begin{figure}[t]
    \centering
    \centerline{\includegraphics[width=\columnwidth, height=6cm, keepaspectratio]{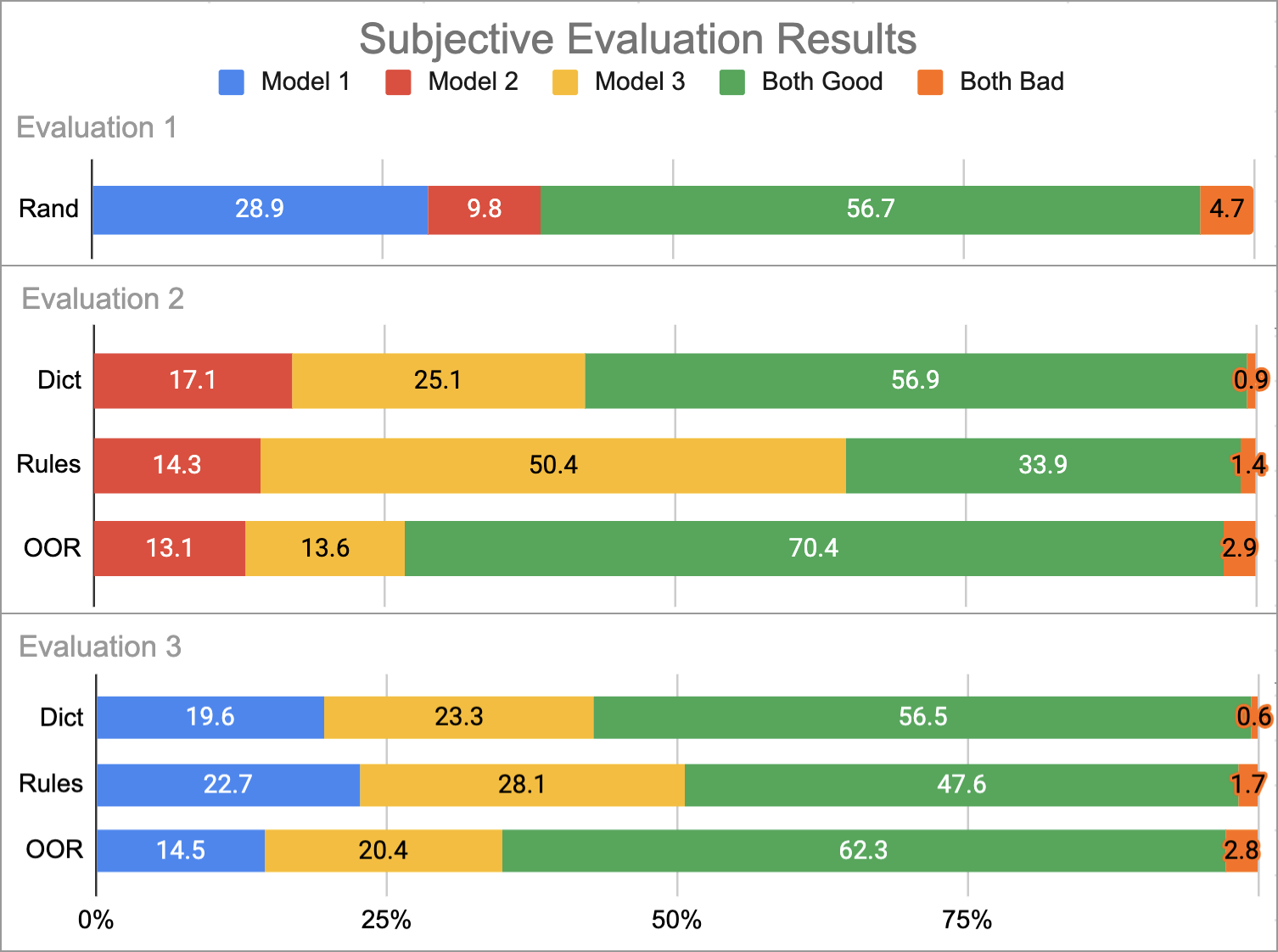}}
    \caption{Results of subjective evaluation indicating the listener's preference. The number indicates the percentage of words in a given evaluation set, falling in one of the categories -- Words for which first model synthesis is preferred; Words for which second model synthesis is preferred; Words for which both the models are good; Words for which both the models are bad.}
    \label{img:mos}
\end{figure}

\section{Results and Discussion}
All the proposed phoneme mapping rules in this paper directly influence the lexicon of only the English words. However, to study if this has affected the modelling of Hindi words during the bilingual TTS system training, we conducted listening tests on randomly chosen 200 Hindi words for the three Evaluations. We observed that the pronunciation was comparable across the three evaluations. 

The summary of the three evaluations conducted on the English words are shown in Figure~\ref{img:mos}.  In evaluation 1, for 28.9\% of words the listeners preferred synthesis using the CMU and CLS phonemes, compared to only 9.8\% of the words for which they preferred synthesis using EngHinCommon phoneset. This suggests that directly mapping individual  CMU phonemes to similar-sounding CLS phonemes does not fix the pronunciation errors in non-Native English lexicon.

The three listening tests of evaluation 2 shown in Figure~\ref{img:mos} prefer pronunciations of the proposed rules (Model 3) over the EngHinCommon phonemes (Model 2). The Model 3 preference is significantly higher for words not part of training vocabulary (`Rules'), achieving an absolute improvement of 35\% over Model 2. This indicates the effectiveness of the proposed rules in correcting the native English lexicon for non-native English speakers. The words in OOR, which had not undergone any of these rules, also improved slightly. This might be a result of reduced confusion between lexicon and pronunciation in the recordings in Model 2. In the Dict words case as well listeners preferred Model 3 over Model 2 (25.1\% vs. 17.1\%). Ideally, since the `dict' words are phonetically corrected for non-native English, Model 3 should have got 100\% preference over \mbox{Model 2}. However, since the lexicon corrections are not speaker specific, the inconsistency in the speaker's pronunciation with the lexicon can be the explanation for this result.
 
In evaluation 3, we compare the proposed phone rules (Model 3) with the baseline phones without any mapping (Model 1). From Figure~\ref{img:mos} it is clear that the listeners preferred the proposed phone rules over the baseline across the three categories of words. On average the proposed rules obtain a 6\% absolute improvement compared to the baseline. 

A statistics of the percentage of the words undergoing each category of the proposed rules are shown in the last column of Table~\ref{tab:rules}. We generated the statistics on the standard CMU dictionary of 130 k unique words. Here we see that around 35\% of the words are getting corrected by the proposed rules (these numbers are excluding the EngHinCommon phoneme mapping). A category-wise statistics are also shown in that table. Further, the actual percentage of training corpus words that get changed depends on the frequency of these dictionary words in the corpus.

\section{Conclusion}
In this work, we have proposed a method to create a non-native English lexicon for Bilingual TTS. We came up with a systematic approach to do triplet analysis which helped uncover inconsistency between native and non-native English lexicons. The proposed method involved limited manual effort in the transliteration of English words from Roman script to Devanagari script. An additional manual effort was required for the analysis of the mismatch between the phoneme sequence of native and non-native pronunciation, to create multiple rules. Formulating these rules based on this mismatch doesn't require a phonetician's expertise. Finally, we showed that proposed non-native lexicon creation helped to improve the synthesis quality of bilingual TTS models with Indian-English and Hindi language pairs. This method can be easily applied to any phonetically orthographic language. Moreover, a reduced phoneset, as used in the proposed method, will improve the speech synthesis in low resource languages. With more data per phoneme, the trained model will be robust. While we have tried to come up with an exhaustive set of rules, some undiscovered rules may still be there, which can be considered as part of future work. The proposed lexicon generation approach can be extended to improve the performance of an Indian English Automatic Speech Recognition system. 

\bibliographystyle{IEEEtran}

\bibliography{mybib}

% Generated by IEEEtran.bst, version: 1.13 (2008/09/30)
\begin{thebibliography}{10}
\providecommand{\url}[1]{#1}
\csname url@samestyle\endcsname
\providecommand{\newblock}{\relax}
\providecommand{\bibinfo}[2]{#2}
\providecommand{\BIBentrySTDinterwordspacing}{\spaceskip=0pt\relax}
\providecommand{\BIBentryALTinterwordstretchfactor}{4}
\providecommand{\BIBentryALTinterwordspacing}{\spaceskip=\fontdimen2\font plus
\BIBentryALTinterwordstretchfactor\fontdimen3\font minus
  \fontdimen4\font\relax}
\providecommand{\BIBforeignlanguage}[2]{{%
\expandafter\ifx\csname l@#1\endcsname\relax
\typeout{** WARNING: IEEEtran.bst: No hyphenation pattern has been}%
\typeout{** loaded for the language `#1'. Using the pattern for}%
\typeout{** the default language instead.}%
\else
\language=\csname l@#1\endcsname
\fi
#2}}
\providecommand{\BIBdecl}{\relax}
\BIBdecl

\bibitem{chu2003microsoft}
M.~Chu, H.~Peng, Y.~Zhao, Z.~Niu, and E.~Chang, ``Microsoft {M}ulan-a bilingual
  {TTS} system,'' in \emph{Proc. {ICASSP}}, vol.~1.\hskip 1em plus 0.5em minus
  0.4em\relax IEEE, 2003, pp. I--I.

\bibitem{lee2018learning}
Y.~Lee, S.~Shon, and T.~Kim, ``Learning pronunciation from a foreign language
  in speech synthesis networks,'' \emph{arXiv preprint arXiv:1811.09364}, 2018.

\bibitem{kumar2007building}
R.~Kumar, R.~Gangadharaiah, S.~Rao, K.~Prahallad, C.~P. Ros{\'e}, and A.~W.
  Black, ``Building a better {I}ndian {E}nglish voice using “more data”,''
  in \emph{Proc. the 6th ISCA workshop on speech synthesis, Germany}, 2007.

\bibitem{yarra2019indic}
C.~Yarra, R.~Aggarwal, A.~Rajpal, and P.~K. Ghosh, ``Indic {TIMIT} and {Indic}
  {English} lexicon: A speech database of {Indian} speakers using {TIMIT}
  stimuli and a lexicon from their mispronunciations,'' in \emph{Proc. Oriental
  COCOSDA}.\hskip 1em plus 0.5em minus 0.4em\relax IEEE, 2019, pp. 1--6.

\bibitem{Thomas2018}
\BIBentryALTinterwordspacing
A.~L. Thomas, A.~Prakash, A.~Baby, and H.~Murthy, ``Code-switching in {Indic}
  speech synthesisers,'' in \emph{Proc. Interspeech}, 2018, pp. 1948--1952.
  [Online]. Available: \url{http://dx.doi.org/10.21437/Interspeech.2018-1178}
\BIBentrySTDinterwordspacing

\bibitem{ramani2013common}
B.~Ramani, S.~L. Christina, G.~A. Rachel, V.~S. Solomi, M.~K. Nandwana,
  A.~Prakash, S.~A. Shanmugam, R.~Krishnan, S.~K. Prahalad, K.~Samudravijaya
  \emph{et~al.}, ``A common attribute based unified {HTS} framework for speech
  synthesis in {I}ndian languages,'' in \emph{ISCA Workshop on Speech
  Synthesis}, 2013.

\bibitem{nlp:tsd16conf}
A.~Baby, N.~L. Nishanthi, A.~L. Thomas, and H.~A. Murthy, ``A unified parser
  for developing {I}ndian language text to speech synthesizers,'' in
  \emph{International Conference on Text, Speech and Dialogue}, 2016.

\bibitem{mahanta2016text}
D.~Mahanta, B.~Sharma, P.~Sarmah, and S.~M. Prasanna, ``{Text to speech
  synthesis system in Indian English},'' in \emph{Proc. {TENCON}}.\hskip 1em
  plus 0.5em minus 0.4em\relax IEEE, 2016, pp. 2614--2618.

\bibitem{saikia2016generating}
R.~Saikia and S.~R. Singh, ``{Generating manipuri english pronunciation
  dictionary using sequence labelling problem},'' in \emph{Proc. International
  Conference on Asian Language Processing}.\hskip 1em plus 0.5em minus
  0.4em\relax IEEE, 2016, pp. 67--70.

\bibitem{babycbblr2016}
A.~Baby, A.~L. Thomas, N.~L. Nishanthi, and T.~Consortium, ``Resources for
  {I}ndian languages,'' in \emph{Community-Based Building of Language
  Resources}, 2016.

\bibitem{vijayarajsolomon2016exploiting}
S.~S. VijayaRajSolomon, V.~Parthasarathy, and N.~Thangavelu, ``Exploiting
  acoustic similarities between {T}amil and {I}ndian {E}nglish in the
  development of an {HMM}-based bilingual synthesiser,'' \emph{IET Signal
  Processing}, vol.~11, no.~3, pp. 332--340, 2016.

\bibitem{maxwell2009acoustic}
O.~Maxwell and J.~Fletcher, ``{Acoustic and durational properties of Indian
  English vowels},'' \emph{World Englishes}, vol.~28, no.~1, pp. 52--69, 2009.

\bibitem{hayashi2019espnettts}
T.~Hayashi, R.~Yamamoto, K.~Inoue, T.~Yoshimura, S.~Watanabe, T.~Toda,
  K.~Takeda, Y.~Zhang, and X.~Tan, ``{ESPnet-TTS}: {U}nified, {R}eproducible,
  and {I}ntegratable open source {E}nd-to-{E}nd {T}ext-to-{S}peech {T}oolkit,''
  \emph{arXiv:1910.10909}, 2019.

\bibitem{shen2018natural}
J.~Shen, R.~Pang, R.~J. Weiss, M.~Schuster, N.~Jaitly, Z.~Yang, Z.~Chen,
  Y.~Zhang, Y.~Wang, R.~Skerrv-Ryan \emph{et~al.}, ``Natural {TTS} synthesis by
  conditioning wavenet on mel spectrogram predictions,'' in \emph{Proc.
  {ICASSP}}.\hskip 1em plus 0.5em minus 0.4em\relax IEEE, 2018.

\bibitem{yamamoto2020parallel}
R.~Yamamoto, E.~Song, and J.-M. Kim, ``Parallel {W}ave{GAN}: {A} fast waveform
  generation model based on generative adversarial networks with
  multi-resolution spectrogram,'' in \emph{Proc. {ICASSP}}.\hskip 1em plus
  0.5em minus 0.4em\relax IEEE, 2020.

\bibitem{oord2016wavenet}
A.~v.~d. Oord, S.~Dieleman, H.~Zen, K.~Simonyan, O.~Vinyals, A.~Graves,
  N.~Kalchbrenner, A.~Senior, and K.~Kavukcuoglu, ``Wavenet: {A} generative
  model for raw audio,'' \emph{arXiv:1609.03499}, 2016.

\bibitem{SIRSA2013393}
\BIBentryALTinterwordspacing
H.~Sirsa and M.~A. Redford, ``The effects of native language on {I}ndian
  {E}nglish sounds and timing patterns,'' \emph{Journal of Phonetics}, vol.~41,
  no.~6, pp. 393--406, 2013. [Online]. Available:
  \url{https://www.sciencedirect.com/science/article/pii/S0095447013000399}
\BIBentrySTDinterwordspacing

\end{thebibliography}

\end{document}